\RequirePackage{lineno}
\documentclass[aps,prd,twocolumn,superscriptaddress,amsmath,amssymb]{revtex4-2}
\def\revtex{true}

\usepackage{etoolbox} 
\usepackage{graphicx} 
\usepackage{epstopdf} 
\usepackage{dcolumn} 
\usepackage{xcolor} 
\usepackage{amsmath,amssymb} 
\usepackage{hyperref} 
\usepackage[utf8]{inputenc} 
\usepackage[english]{babel} 
\usepackage{ifthen} 
\usepackage{blindtext} 
\usepackage{orcidlink} 
\usepackage{datetime2} 
\ifthenelse{\isundefined{\revtex}}{\usepackage[displaymath, mathlines]{lineno}}{} 

\graphicspath{{figures/}} 

\newboolean{articletitles}
\setboolean{articletitles}{true} 

\RequirePackage{xspace}
\RequirePackage{relsize}

\def\babar{\mbox{\slshape B\kern-0.1em{\smaller A}\kern-0.1em
    B\kern-0.1em{\smaller A\kern-0.2em R}}\xspace}

\mathchardef\PDelta="7101
\mathchardef\PXi="7104
\mathchardef\PLambda="7103
\mathchardef\PSigma="7106
\mathchardef\POmega="710A
\mathchardef\PUpsilon="7107
                 
\def\PB      {\ensuremath{B}\xspace}                 
                 
\def\PD      {\ensuremath{D}\xspace}

\def\PK      {\ensuremath{K}\xspace}

\def\kaon  {\ensuremath{\PK}\xspace}
\def\Kbar  {\kern 0.2em\overline{\kern -0.2em \PK}{}\xspace}

\def\Kz    {\ensuremath{\kaon^0}\xspace}
\def\Kzb   {\ensuremath{\Kbar^0}\xspace}
\def\KzKzb {\ensuremath{\Kz \kern -0.16em \Kzb}\xspace}
\def\Kp    {\ensuremath{\kaon^+}\xspace}
\def\Km    {\ensuremath{\kaon^-}\xspace}

\def\KpKm  {\ensuremath{\Kp \kern -0.16em \Km}\xspace}

\def\D       {\ensuremath{\PD}\xspace}
\def\Dbar    {\kern 0.2em\overline{\kern -0.2em \PD}{}\xspace}

\def\Dz      {\ensuremath{\D^0}\xspace}
\def\Dzb     {\ensuremath{\Dbar^0}\xspace}
\def\DzDzb   {\ensuremath{\Dz {\kern -0.16em \Dzb}}\xspace}
\def\Dp      {\ensuremath{\D^+}\xspace}
\def\Dm      {\ensuremath{\D^-}\xspace}

\def\DpDm    {\ensuremath{\Dp {\kern -0.16em \Dm}}\xspace}

\def\Bbar    {\ensuremath{\kern 0.18em\overline{\kern -0.18em \PB}{}}\xspace}

\def\Y#1S{\ensuremath{\PUpsilon{(#1S)}}\xspace}

\def\Lbar {\ensuremath{\kern 0.1em\overline{\kern -0.1em\PLambda}}\xspace}

\def\to                 {\ensuremath{\rightarrow}\xspace}

\def\CP                {\ensuremath{C\!P}\xspace}

\newcommand{\tev}{\ensuremath{\mathrm{\,Te\kern -0.1em V}}\xspace}
\newcommand{\gev}{\ensuremath{\mathrm{\,Ge\kern -0.1em V}}\xspace}
\newcommand{\mev}{\ensuremath{\mathrm{\,Me\kern -0.1em V}}\xspace}
\newcommand{\kev}{\ensuremath{\mathrm{\,ke\kern -0.1em V}}\xspace}
\newcommand{\ev}{\ensuremath{\mathrm{\,e\kern -0.1em V}}\xspace}
\newcommand{\gevc}{\ensuremath{{\mathrm{\,Ge\kern -0.1em V\!/}c}}\xspace}
\newcommand{\mevc}{\ensuremath{{\mathrm{\,Me\kern -0.1em V\!/}c}}\xspace}
\newcommand{\gevcc}{\ensuremath{{\mathrm{\,Ge\kern -0.1em V\!/}c^2}}\xspace}
\newcommand{\gevgevcccc}{\ensuremath{{\mathrm{\,Ge\kern -0.1em V^2\!/}c^4}}\xspace}
\newcommand{\mevcc}{\ensuremath{{\mathrm{\,Me\kern -0.1em V\!/}c^2}}\xspace}

\def\gsim{{~\raise.15em\hbox{$>$}\kern-.85em
          \lower.35em\hbox{$\sim$}~}\xspace}
\def\lsim{{~\raise.15em\hbox{$<$}\kern-.85em
          \lower.35em\hbox{$\sim$}~}\xspace}

\def\simge{\mathrel{%
   \rlap{\raise 0.511ex \hbox{$>$}}{\lower 0.511ex \hbox{$\sim$}}}}
\def\simle{\mathrel{
   \rlap{\raise 0.511ex \hbox{$<$}}{\lower 0.511ex \hbox{$\sim$}}}}

\begin{document}

\begin{flushright}
\begin{tabular}{l}
Belle II Preprint 2026-022 \\
KEK Preprint 2026-19 
\end{tabular}
\end{flushright}

\title{Measurement of the time-dependent \CP\ asymmetry in $B^0\to\rho^0\gamma$ decays
}

  \author{M.~Abumusabh\,\orcidlink{0009-0004-1031-5425}} 
  \author{I.~Adachi\,\orcidlink{0000-0003-2287-0173}} 
  \author{A.~Aggarwal\,\orcidlink{0000-0002-5623-3896}} 
  \author{H.~Ahmed\,\orcidlink{0000-0003-3976-7498}} 
  \author{Y.~Ahn\,\orcidlink{0000-0001-6820-0576}} 
  \author{H.~Aihara\,\orcidlink{0000-0002-1907-5964}} 
  \author{M.~Akdag\,\orcidlink{0009-0004-3728-1077}} 
  \author{N.~Akopov\,\orcidlink{0000-0002-4425-2096}} 
  \author{A.~Akram\,\orcidlink{0000-0003-0198-5852}} 
  \author{S.~Alghamdi\,\orcidlink{0000-0001-7609-112X}} 
% ...
\collaboration{The Belle and Belle II Collaborations}

\begin{abstract}
We present a measurement of the time-dependent \CP\ asymmetry in $B^{0}\rightarrow\rho^{0}\gamma$ 
decays using a large data sample of $1.29\times 10^9$ $B\overline{B}$ pairs collected by the 
Belle and Belle~II experiments in $e^{+}e^{-}$ collisions at the $\Upsilon (4S)$ resonance.
We perform a simultaneous fit to the Belle and Belle~II datasets and measure \CP-violating 
parameters $C_{\rho\gamma}=0.59\pm 0.23\pm 0.05$ and $S_{\rho\gamma}=-0.86\pm 0.51\pm 0.12$.
These are the most precise measurements to date of \CP\ violation in this radiative
penguin decay. The measured values are consistent with expectations based on the Standard Model.

\end{abstract}

\maketitle

\section{Introduction}
Flavor-changing neutral-current (FCNC) processes, especially the
$b\to (s,d)\gamma$ transition, are sensitive probes of physics beyond the 
Standard Model (SM)~\cite{Glashow:1970gm,Arhrib:2001gr,Ali:2002kw,Xiao:2003vq,Li:2013vlx,Kou:2018nap}.
Due to the $V\!-\!A$ structure of the weak interaction, the daughter $(s,d)$ quark is left-handed. 
This corresponds to its spin being aligned opposite to its momentum, and the resulting 
$\overline{K}{}^*$ or $\rho$ meson must have a spin component $s^{}_z\!=\!0$ or 
$s^{}_z\!=\!-1$ along the decay axis~$z$. 
As the final state has $J\!=\!0$ and the $\overline{K}{}^*$ or $\rho$
recoils against a photon, $s^{}_z\!=\!0$ is excluded. Thus, the vector meson
must have $s^{}_z\!=\!-1$, and to conserve angular momentum the photon must have negative 
helicity. For the photon to emerge with positive helicity, the chirality of the $(s,d)$ 
quark must be flipped; this introduces a suppression factor
of $m^{}_{s,d}/m^{}_{b}\simle 0.07$~\cite{Beneke:2004dp}.
For the charge-conjugate $\bar{b}\to (\bar{s},\bar{d})\gamma$ transition,
the photon must have positive helicity;
thus, the final states for $\overline{B}{}^0\to\rho^0\gamma^{}_{-}$ 
and $\overline{B}{}^0\to B^0\to\rho^0\gamma^{}_{+}$ are distinct, and
without a chirality flip the amplitudes do not interfere. Consequently, 
\CP\ violation arising from mixing (indirect \CP\ violation) 
is expected to be very small, 
and measuring a significant \CP\ asymmetry would be 
interpreted as evidence for physics beyond the~SM.
One extension of the SM, ``left-right symmetric'' models (LRSM), 
can lift the helicity suppression and give rise to 
\CP\ violation at the few percent level~\cite{Ball:2006eu}.

Experimentally, there have been several searches for \CP\ violation in 
$B^0\to K^{*0}\gamma$ decays~\cite{Belle:2006pxp,BaBar:2008okc,Belle-II:2024uzp},
and one search in $B^0\to\rho^0\gamma$ decays~\cite{Belle:2007kjy}.
The latter was performed by the Belle experiment using $0.66\times 10^9$
$B\overline{B}$ pairs. Here we update this measurement using twice as much
data: $0.77\times 10^9$ $B\overline{B}$ pairs from the complete data set
of Belle, and an additional $0.52\times 10^9$ $B\overline{B}$ pairs from Belle~II. 

In Belle and Belle~II, $B\overline{B}$ 
pairs are produced in a quantum-correlated state.
One of the $B$'s is referred to as the ``tagging'' $B$ ($B_{\rm tag}$), while the
other is referred to as the signal $B$  ($B_{\rm sig}$). At the moment $B_{\rm tag}$ 
decays into a flavor-specific final state, $B_{\rm sig}$ is projected into a state of 
the opposite flavor. In this manner, $B_{\rm tag}$ identifies the initial flavor of 
$B_{\rm sig}$. The time-dependent {\it CP} asymmetry of $B^0\to\rho^0\gamma$
decays is defined as
\begin{equation}
 A_{CP}(\Delta t) = \frac{\Gamma\left(\overline{B}{}^0(\Delta t)\to\rho^{0}\gamma\right)
-\Gamma\left(B^0(\Delta t)\to\rho^{0}\gamma\right)}
{\Gamma\left(\overline{B}{}^0(\Delta t)\to\rho^{0}\gamma\right) + 
\Gamma\left(B^0(\Delta t)\to\rho^{0}\gamma\right)}\,,
\end{equation}
where $\Gamma(B\to\!f)$ denotes the partial width and $\Delta t$ is 
the difference in proper decay times between $B_{\rm sig}$ and $B_{\rm tag}$. 
This asymmetry can be parametrized as
\begin{equation}
 A_{\it CP}(\Delta t) \,=\,
- C^{}_{\rho\gamma}\cos(\Delta m\Delta t) \,+\, S^{}_{\rho\gamma}\sin(\Delta m\Delta t) \,,
\label{eqn:ACP}
\end{equation}
where $C_{\rho\gamma}$ and $S_{\rho\gamma}$ arise from direct \CP\ violation and 
indirect \CP\ violation, respectively, and $\Delta m$ is the mass difference 
between the two mass eigenstates of the $B^0$-$\overline{B}{}^0$ system. 
The Standard Model (SM) expectations are 
$C_{\rho\gamma}\!= 0.17\,^{+0.07}_{-0.05}$~\cite{Beneke:2004dp}~\footnote{The SM prediction is for the 
time-integrated asymmetry: $A^{}_{CP}\!=\!(-10.4\,^{+3.4}_{-4.3})\%$~\cite{Beneke:2004dp}. To obtain 
$C^{}_{\rho^0\gamma}$, we insert this into the relation $A^{}_{CP}\!=\!-C^{}_{\rho^0\gamma}/(1+x_d^2)$, 
where $x^{}_d$ is the $B^0$-$\overline{B}{}^0$ mixing parameter.}
and $S_{\rho\gamma}\!< 0.02$~\cite{Ball:2006eu}.
Previously, Belle measured $C_{\rho\gamma}=0.44\pm 0.49 \pm 0.14$ 
and $S_{\rho\gamma}=-0.83\pm 0.65\pm 0.18$~\cite{Belle:2007kjy}.

This paper is organized as follows.
The Belle and Belle~II detectors and datasets used are described in Sec.~\ref{sec:setup}.
Event reconstruction and selection criteria are explained in Sec.~\ref{sec:selection}.
Additional criteria to suppress backgrounds are described in Sec.~\ref{sec:back_suppression}.
The fitting procedure used to extract the signal yield is discussed in Sec.~\ref{sec:yield_fit},
and the fitting procedure used to measure the \CP\ asymmetries is discussed in Sec.~\ref{sec:cp_fit}.
Systematic uncertainties are described in Sec.~\ref{sec:systematics}, and we conclude in 
Sec.~\ref{sec:conclusion}.

\section{Detectors and datasets}
\label{sec:setup}

The Belle detector~\cite{Belle:2000cnh,Brodzicka:2012jm} was a large-solid-angle spectrometer that operated 
at the KEKB asymmetric-energy $e^{+}e^{-}$ (3.5~GeV on 8.0~GeV) collider~\cite{Kurokawa:2001nw,Abe:2013kxa} 
from 1999 to 2010. The detector consisted of a four-layer
silicon vertex detector and a central drift chamber (CDC) 
for reconstructing trajectories of charged particles (tracks), an array of aerogel Cherenkov counters (ACC) 
and time-of-flight scintillation counters (TOF) for identifying charged particles, and an electromagnetic 
CsI(Tl) crystal calorimeter (ECL) for identifying photons and electrons. These detectors were surrounded 
by a superconducting solenoid coil providing a magnetic field of 1.5~T. An iron flux return yoke located 
outside the coil was instrumented with resistive-plate chambers to detect $K^{0}_{L}$ mesons and identify 
muons (KLM).

The Belle~II detector~\cite{Abe:2010gxa} is an upgrade of the Belle detector 
and operates at the SuperKEKB $e^{+}e^{-}$ collider. The energies of the positron 
and electron beams are 4.0~GeV and 7.0~GeV,
respectively. The vertex detector consists of two parts: 
a two-layer silicon-pixel detector (PXD) surrounded by 
a four-layer double-sided silicon-strip detector.
The Belle~II CDC is radially larger than that 
for Belle and surrounded by two types of Cherenkov light detector systems used for particle 
identification: a time-of-propagation detector (TOP) in the barrel region (32.2$^{\circ}$ to 
128.7$^{\circ}$), and an aerogel ring-imaging Cherenkov detector (ARICH) in the forward endcap 
region (12.4$^{\circ}$ to 31.4$^{\circ}$). The Belle~II ECL is re-used (with improved electronics)
from Belle along with the solenoid magnet and the iron flux return yoke. Part of the Belle~II KLM 
is instrumented with scintillator strips instead of resistive-plate chambers to reduce 
deadtime. For both Belle and Belle~II, the $z$ axis is defined as the central axis of 
the solenoid, with the positive direction being very close to the direction of the 
electron beam.

Both Belle and Belle~II datasets were recorded at an $e^+e^-$ center-of-mass (c.m.) energy 
corresponding to the $\Upsilon(4S)$ resonance. The Belle dataset consists of the full 
711~$\mathrm {fb}^{-1}$ recorded at this energy. The Belle~II dataset consists of two parts: 
405~$\mathrm {fb}^{-1}$ that included the PXD, and 88~$\mathrm {fb}^{-1}$ in which the 
PXD was turned off due to the risk of damage from beam loss. 
For the first 365~$\mathrm {fb}^{-1}$ of data with the PXD, only one sixth of the 
second layer was installed. We measure \CP\ parameters $C$ and $S$ by performing 
a simultaneous fit to all three datasets.

Monte Carlo (MC) simulated samples are used to optimize selection criteria, study 
sources of backgrounds, calculate reconstruction efficiencies, and determine probability 
density functions used for fitting the data. We use {\sc EvtGen}~\cite{Lange:2001uf} 
to generate $e^{+}e^{-}\to B\overline{B}$ events and simulate particle decays.
For the Belle~II sample, we also employ {\sc Pythia} 8.2~\cite{Sjostrand:2014zea}.
We simulate $e^+e^-\to q\bar{q}~(q=u,d,s,c)$ continuum background events using 
{\sc Pythia} 6.4~\cite{Sjostrand:2006za} for Belle, and KKMC~\cite{Ping:2008zz} 
for hard scattering followed by {\sc Pythia} 8.2~\cite{Sjostrand:2014zea} for 
hadronization for Belle~II.
We simulate final-state radiation using {\sc Photos}~\cite{Barberio:1990ms,Barberio:1993qi} 
for both Belle and Belle~II.
The detector response is simulated with {\sc geant3}~\cite{Brun:1987ma} 
for Belle and {\sc geant4}~\cite{Agostinelli:2002hh} for Belle~II.
All data and simulated samples are processed using the Belle~II analysis 
software framework~\cite{Kuhr:2018lps,basf2-zenodo,b2bii}.

\section{Event selection}
\label{sec:selection}

The selection criteria used for Belle and Belle~II data are very similar.
However, certain criteria are adjusted to account for the improved performance 
of the Belle~II detector.

The online triggers for Belle and Belle~II are based on either the number of tracks, 
the number of ECL energy deposits (clusters), or the total energy deposited in all 
clusters. For both experiments, the trigger efficiency for signal events is almost 100\%.

In the offline analysis, a high energy primary photon candidate ($\gamma_{\rm prim}$) 
is reconstructed from an ECL cluster not associated with any track. Only ECL clusters 
whose polar angle $\theta$ with respect to the $+z$ axis is in the barrel region are 
considered. The energy of $\gamma_{\rm prim}$ in the $e^+e^-$ c.m.\ frame, denoted 
$E^{\ast}_{\gamma}$, is required to be in the range $(1.8, 3.0)~{\rm GeV}$~\footnote{Throughout 
this paper, quantities with asterisks are calculated in the $e^{+}e^{-}$ c.m. frame.}.
For Belle, we require $E_{\rm 9}/E_{\rm 25}>0.95$, where $E_{\rm 9}/E_{\rm 25}$ is the ratio of
energies deposited within a $3\times 3$ array of CsI(Tl) crystals and within a $5\times 5$ 
array, respectively, both centered on the crystal with the highest energy. This requirement 
selects cluster shapes consistent with an electromagnetic shower. For Belle~II, we require 
$E_{\rm 9}/E_{\rm 21}>0.95$, where $E_{21}$ is defined similarly to $E_{25}$ but with energy 
deposits at the four corners of the $5\times 5$ array removed. 

We also define for Belle~II a cluster second moment 
$\xi\equiv \sum_{i} \bigl(\alpha^{}_{i}E_{i}r_{i}^{2}\bigr) / \sum_{i} \bigl(\alpha^{}_{i}E_{i}\bigr)$,
where $E_{i}$ is the energy deposited in the $i$th crystal of a cluster;
$r_{i}$ is the distance in the plane perpendicular to the shower axis
from the center of the cluster to the center of the $i$th 
crystal~\footnote{The shower axis is defined by the line from the $e^+e^-$ 
collision point to the energy-weighted centroid of the ECL cluster.};
$\alpha^{}_i$ is a weighting factor ranging from $0\!-\!1$ to allow for $E^{}_i$ to be 
shared among more than one cluster; and the sum is taken over all crystals in the cluster. 
To reject energy deposits not resulting from electromagnetic showers, we require that 
there be at least eight crystals with deposited energy in a cluster. To better identify 
the primary photon cluster, we require $\xi<1.5$~cm$^2$.

Track candidates are required to satisfy $dr<0.5$~cm and $|dz|<2$~cm,
where $dr$ and $dz$ are transverse and longitudinal impact parameters, respectively.
We select $\pi^{+}$ candidates by requiring that a track satisfy 
${\cal R}^{}_{\pi}\!\equiv\!{\cal L}^{}_{\pi} / ({\cal L}^{}_{\pi} + {\cal L}^{}_K )>0.90$,
where ${\cal L}^{}_{\pi (K)}$ is the likelihood for the pion (kaon) hypothesis.
These likelihoods are based on information from the CDC, ACC, and TOF 
detectors in Belle~\cite{Nakano:2002jw}, and from the CDC, TOP, and ARICH 
detectors in Belle~II~\cite{Belle-II:2025tpe}.
This requirement is $>\!85$\% efficient for pions from signal decays and 
rejects $>\!95$\% of kaons from $B^0\to K^{*0}(\to K^+\pi^-)\gamma$ decays.

We reconstruct $\rho^{0}\rightarrow\pi^{+}\pi^{-}$ candidates by requiring that 
the invariant mass of the pion pair $M(\pi^+\pi^-)$
be in the range ${\rm [0.64, 0.90]~GeV}/c^{2}$ 
for Belle data, and in the range ${\rm [0.65, 0.90]~GeV}/c^{2}$ for Belle~II data.
The slight difference is due to a slight shift in the $M(\pi^{+}\pi^{-})$ spectrum 
for Belle~II. A $B$ candidate is subsequently reconstructed by combining a $\rho^0$ 
candidate with a $\gamma_{\rm prim}$ candidate. To remove 
$e^+e^-\!\!\to\!q\overline{q}$ events, we require that there be at least two 
additional tracks in the event and that the ratio of the second to zeroth 
Fox-Wolfram moments ($R^{}_2$)~\cite{Fox:1978vu} be less than~0.7.

To fit for the signal yield, we define two observables:
the beam-energy-constrained mass 
$M_{\rm bc}\equiv\sqrt{\left(E_{\rm beam}^{\ast}/c^{2}\right)^2 - |{\bf p}_{B}^{\ast}/c|^2}$,
and the energy difference $\Delta E\equiv E_{B}^{\ast}-E_{\rm beam}^{\ast}$, where 
$E^\ast_{\rm beam}$ is the beam energy and $E^\ast_{B}$ and $p^\ast_{B}$ are the 
reconstructed energy and momentum, respectively, of the signal $B$ candidate.
All quantities are calculated in the $e^+e^-$ c.m.\ frame.
To improve the resolution in $M_{\rm bc}$, we substitute for the magnitude of the photon 
momentum the difference between the beam energy and the energy of the $\rho^{0}$ candidate:
$\mbox{\boldmath $\vec{p}$}^{\,\ast}_{B^0}\rightarrow \mbox{\boldmath $\vec{p}$}^{\,\ast}_{\rho^{0}} 
+ \left(\mbox{\boldmath $\vec{p}$}^{\,\ast}_{\gamma} / 
|\mbox{\boldmath $\vec{p}$}^{\,\ast}_{\gamma}|\right) \left(E^{\ast}_{\rm beam}-E^{\ast}_{\rho^{0}}\right)/c$,
where $E_{\rho^{0}}^{\ast}$ and $\mbox{\boldmath $\vec{p}$}^{\,\ast}_{\rho^{0}}$ are the energy and momentum, 
respectively, of the $\rho^{0}$ candidate, and $\mbox{\boldmath $\vec{p}$}^{\,\ast}_{\gamma}$ is 
the momentum of the $\gamma_{\rm prim}$ candidate.
According to MC simulation, this modification reduces the width of the 
$M_{\rm bc}$ distribution by about~10\%.
We retain signal candidates that satisfy 
$M_{\rm bc}>5.24~{\rm GeV/}c^{2}$ and $|\Delta E|<0.3~{\rm GeV}$.

In the $\Delta t$ fit to determine parameters
$C^{}_{\rho\gamma}$ and $S^{}_{\rho\gamma}$, the most problematic background is 
$B^0\to K^{\ast}(892)^0\gamma$, which is kinematically similar to $B^0\to\rho^0\gamma$
and has a branching fraction approximately fifty times larger. To better discriminate 
against this backgound, we define a variable $M(K\pi)$, which is the invariant mass of 
the $\rho^{0}$ candidate as calculated assuming the charged pion with the lower value of 
${\cal R}^{}_{\pi/K}$ is a kaon. This allows us to distinguish $B^{0}\rightarrow\rho^{0}\gamma$ 
from $B^{0}\rightarrow K^{\ast 0}\gamma$, which peaks in the $K^{\ast}\left(892\right)^0$ mass region.
We require $M(K\pi)\!\in\!(0.92, 1.50)~{\rm GeV}/c^{2}$ and use $M(K\pi)$ in the 
fitting procedure to better discriminate $B\rightarrow K^{\ast}\gamma$ decays from the signal.

To reduce background arising from other (misreconstructed) $B\overline{B}$ events, the $B$ 
candidate must satisfy $|\cos{\theta^{}_\gamma}|<0.8$, where $\theta^{}_\gamma$ is the angle in the 
$B$ rest frame between the momentum of $\gamma_{\rm prim}$ and the opposite of the boost direction 
of the laboratory frame. For correctly reconstructed signal decays, the $|\cos{\theta^{}_\gamma}|$ 
distribution is uniform, whereas for $B\overline{B}$ background the distribution tends to peak 
at $|\cos{\theta^{}_\gamma}|\approx 1$.
After applying the selection criteria listed above, 3.0\% of events have multiple 
$B^{0}\rightarrow\rho^{0}\gamma$ candidates. For such multiple candidate events, 
the average multiplicity is 2.3. To minimize potential bias, we select one candidate randomly.

After the reconstruction of $B_{\rm sig}$, the remaining particles 
in the ``rest of event'' (ROE) are assumed to originate from the $B_{\rm tag}$ decay.
The flavor of $B_{\rm tag}$ is determined from the properties of ROE particles
by a flavor-tagging algorithm based on a graph neural network (GNN)~\cite{Belle-II:2024lwr}.

To assess the quality of the simulation, we compare various MC distributions with 
corresponding data distributions from four control samples:
$B^+\to \overline{D}{}^0 (\to\!K^+\pi^-)\,\pi^{+}$, 
$B^0\to D^- (\to\!K^+\pi^-\pi^-)\,\pi^{+}$, 
$B^0\to K^{\ast 0} (\to\!K^+\pi^-)\,\gamma$, 
and $B^{0}\to J/\psi\,K^{\ast 0} (\to\!K^+\pi^-)$.
These control sample studies are discussed in the following sections.
For the $\overline{D}{}^0$ and $D^-$ decays, we require that
the $K^+$ and $\pi^\pm$ daughters satisfy 
${\cal R}^{}_{K}\!>\!0.6$ and ${\cal R}^{}_{\pi}\!>\!0.6$, respectively,
and that the invariant masses satisfy
${\rm 1.85~GeV}/c^{2}<M(K^{+}\pi^{-})<{\rm 1.88~GeV}/c^{2}$ and 
${\rm 1.86~GeV}/c^{2}<M(K^{+}\pi^{-}\pi^{-})<{\rm 1.88~GeV}/c^{2}$.
For the $K^{\ast 0}$ decays, we require that the charged kaon have
${\cal R}^{}_{K}\!>\!0.6$ and that the 
$\pi^{-}$ satisfy the same criteria as applied to signal 
$B^{0}\to\rho^{0}\gamma$ decays. The invariant mass of the 
$K^+\pi^-$ pair must satisfy 
${\rm 0.817~GeV}/c^{2}\!<\!M(K^{+}\pi^{-})\!<\!{\rm 0.967~GeV}/c^2$.
The $J/\psi$ mesons are reconstructed via $J/\psi\rightarrow\mu^{+}\mu^{-}$ 
and $J/\psi\rightarrow e^{+} e^{-}$, where muon and electron candidates
are identified using information mainly from the KLM and ECL detectors, 
respectively. This information is used to calculate likelihoods ${\cal L}^{}_\xi$ 
for different particle types $\xi$~\cite{Hanagaki:2001fz,Abashian:2002bd,Belle-II:2026okl}, 
and the lepton candidate $\ell^\pm$ is selected by requiring that the ratio
${\cal R}^{}_{\ell} \equiv {\cal L}^{}_\ell
/({\cal L}^{}_{\mu}+{\cal L}^{}_{e}+{\cal L}^{}_{\pi}+{\cal L}^{}_{K}+{\cal L}^{}_{p})$
be greater than 0.90. The invariant mass of the $J/\psi$ candidate must satisfy 
${\rm 2.95~GeV}/c^{2}<M_{J/\psi}<{\rm 3.15~GeV}/c^{2}$.

\section{Background suppression}
\label{sec:back_suppression}

After applying the above selection criteria, the remaining background is 
mainly from $q\overline{q}$ events, where the candidate $\gamma_{\rm prim}$ 
results from an asymmetric $\pi^{0}\rightarrow\gamma\gamma$ decay. 
In this case, one of the photons has much higher energy than the other.
To reduce this background, we use two boosted decision trees:
a dedicated $\pi^{0}$ veto and one for $q\overline{q}$ suppression.

\subsection{$\pi^{0}$ veto}
\label{sec:pi0etaveto}
For the $\pi^{0}$ veto, we pair $\gamma_{\rm prim}$ with other 
photons in the event to reconstruct possible $\pi^{0}\to\gamma\gamma$ decays.
If such a candidate is identified, the $\gamma_{\rm prim}$ candidate is removed.
The energy of the photon with which $\gamma_{\rm prim}$ is paired ($\gamma_{\rm soft}$) 
must exceed a polar-angle-dependent threshold. For Belle~II data,
we require that the sum of crystal weights $w_{i}$ for the $\gamma_{\rm soft}$ cluster be 
at least two, and that this cluster be reconstructed in time with the event trigger.
The probability of a correctly reconstructed $\pi^{0}$ is calculated using a boosted 
decision tree ($BDT_{v}$) trained on simulated events. 
The variables used in $BDT_{v}$ are $R^{}_2$; 
the cosine of the angle between the thrust axis~\cite{Farhi:1977sg} 
of the signal $B$ and the $z$ axis;
the cosine of the helicity angle of the $B^0\to\rho^0\gamma$ decay;
the cosine of the angle between the thrust axis of the signal $B$ and the 
thrust axis of the rest of the event ($\cos{\theta^*_{T_{B}T_{O}}}$);
several modified Fox-Wolfram moments~\cite{Fox:1978vu,Belle:2003fgr}; and 
the sphericity of ROE particles in the event~\cite{Bjorken:1969wi}.
All these quantitities are calculated in the $e^+e^-$ c.m.\ frame.
To reject $\gamma_{\rm prim}$ candidates resulting from $\pi^0\to\gamma\gamma$ decays, 
we require that the $BDT_{v}$ output be less than 0.85. This value is found from 
MC simulation to maximize the precision of the fitted \CP\ parameters. It reduces 
background from $B\to\rho\pi^0$ to only 2--3\% of all $B\overline{B}$ background.

The signal efficiency is calculated from simulation and corrected for 
possible differences between data and simulation using the ratio of efficiencies 
measured in data and simulation for $B\to\overline{D}\pi^{+}$ control samples,
where $\overline{D}{}^0\to K^+\pi^-$ and 
$D^{-}\rightarrow K^+\pi^-\pi^-$~\footnote{Cabibbo-favored 
all-charged $B\to \overline{D}\pi^+$ decays are used, as background is 
negligible and photon reconstruction does not introduce a \CP\ asymmetry.}.
In this procedure, the $\pi^+$ from the $B$ decay is treated as a prompt 
photon, analogous to that from signal $B^0\to\rho^0\gamma$ decays.
The resulting efficiencies are 
$0.936\pm 0.001$ for Belle and 
$0.939\pm 0.001$ for Belle~II, 
and the respective correction factors are 
$0.976\pm 0.008$ and $1.058\pm 0.012$.
More details are given in Ref.~\cite{Belle:2024qhe}.

\subsection{$q\overline{q}$ suppression}
\label{sec:qqsup}

To further reduce contamination from $q\bar{q}$ background, we introduce another boosted 
decision tree classifier ($BDT_{q\overline{q}}$) that relies on differences in event shapes 
between $q\overline{q}$ and $B\overline{B}$ events.
The variables used in $BDT_{q\overline{q}}$ are 
$\cos{\theta^*_{T_{B}T_{O}}}$~\cite{Farhi:1977sg};
several modified Fox-Wolfram moments~\cite{Fox:1978vu,Belle:2003fgr};
the sphericity of ROE particles in the event~\cite{Bjorken:1969wi};
and the cosine of the polar angle of the signal~$B$.
The $\cos{\theta^*_{T_{B}T_{O}}}$ variable is the most discriminating.
The $BDT_{q\overline{q}}$ training uses simulated $B\to\rho\gamma$ decays 
and continuum events, and the training is performed separately for 
Belle and Belle~II data.
Similar to the $\pi^{0}$ veto, the requirement on the $BDT_{q\overline{q}}$ 
output ($>\!0.75$) is chosen to maximize the sensitivity to the \CP\ 
parameters, as found from MC simulation studies. It reduces $q\bar{q}$ 
background by a factor of 16 for Belle and 18 for Belle~II.

The signal efficiencies are calculated from simulation and corrected for possible 
differences between data and simulation using the $B\to K^{\ast}\gamma$ control samples.
These efficiencies are $0.704\pm 0.001$ for Belle and $0.756\pm 0.001$ for Belle~II,
and the respective correction factors are $0.967\pm 0.019$ and $0.978\pm 0.018$.

\section{Fitting for the signal yield}
\label{sec:yield_fit}

To determine the signal yield, we perform an extended unbinned maximum-likelihood 
fit to the observables $M_{bc}$, $\Delta E$, and $M(K\pi)$. 
The fit is performed simultaneously to the three separate datasets:
Belle, Belle~II with the PXD, and Belle~II without the PXD.
The Belle~II dataset is divided into two parts due to a 
difference in the decay-time resolution.
In the fit we model four components: $B\to\rho\gamma$ signal decays, 
$B\to K^{\ast}\gamma$ decays, other $B\overline{B}$ backgrounds, and $q\bar{q}$ background.
A small contribution from nonresonant $B^0\to\pi^+\pi^-\gamma$ 
decays is treated as part of the signal component.
We fix the ratios of signal yields among the three datasets according 
to their luminosities and reconstruction efficiencies; this is equivalent 
to fitting for a common $B^0\to\rho^0\gamma$ branching fraction.

As correlations among the fitted observables are very small, the overall probability
density function (PDF) for each component is taken as the product of PDFs for the
individual observables. These PDFs are as follows.
For signal, we model both the $M_{\rm bc}$ and $\Delta E$ distributions with 
Crystal Ball functions~\cite{Oreglia:1981fx}, and the $M(K\pi)$ distribution 
with a Novosibirsk function~\cite{Belle:1999bhb} convolved with a Gaussian.
For $B\rightarrow K^{\ast}\gamma$ decays, a double-sided Crystal Ball function 
is used to model $M_{bc}$, a regular Crystal Ball function is used for $\Delta E$, 
and kernel density estimation is used for $M(K\pi)$. 
The shape of the kernel density is determined from MC simulation. For $B\overline{B}$ background,
we use the sum of a Crystal Ball function and an ARGUS function~\cite{ARGUS:1990hfq} for $M_{\rm bc}$,
an exponential function for $\Delta E$, and kernel density estimation for $M(K\pi)$.
Finally, for $q\overline{q}$ background, we use an ARGUS function for $M_{\rm bc}$, a quadratic 
Chebychev polynomial for $\Delta E$, and kernel density estimation for $M(K\pi)$.

The shape parameters of the $\Delta E$ distribution for $q\overline{q}$ background are 
floated in the fit, while all other $q\bar{q}$ shape parameters are fixed to values 
obtained from MC simulation. These values are obtained separately for each data set.
To account for possible differences between data and simulation, the 
$M_{\rm bc}$ and $\Delta E$ distributions for signal decays are corrected for small differences 
observed between data and simulation for the $B\to K^{\ast}\gamma$ control samples.
Similarly, the parameters for the $M_{\rm bc}$ distribution of $q\overline{q}$ background 
are corrected according to small differences observed between data and simulation in the 
sideband region $BDT_{q\overline{q}}\in [0.40,0.75]$.
The kernel density estimate for the $M(K\pi)$ distribution of $q\overline{q}$ 
background is determined from the $BDT_{q\overline{q}}$ sideband.

The yields of $B^{0}\to K^{\ast 0}\gamma$ and other $B\overline{B}$ backgrounds are fixed 
in the fit, whereas the yields of signal and $q\overline{q}$ background are floated.
The $K^{\ast}\gamma$ background yield is estimated as 
$N_{K^{\ast}\gamma}\!=\!2\cdot N_{\Upsilon(4S)}\cdot f_{00}\cdot 
{\cal B}(B\to K^{\ast}\gamma)\cdot F_{K/\pi}\cdot\varepsilon^{}_{K^{\ast}\gamma}$,
where $N_{\Upsilon(4S)}$ is the number of $\Upsilon(4S)\to B\overline{B}$ decays,
$f_{00}$ is the fraction of $\Upsilon (4S)$ decays that produce $B^{0}\overline{B}{}^{0}$,
${\cal B}(B\to K^{\ast}\gamma)$ is the branching fraction,
$F_{K/\pi}$ is the probability for a charged kaon to be misidentified as a pion, and
$\varepsilon^{}_{K^{\ast}\gamma}$ is the efficiency to select $K^{\ast}\gamma$ events.
The $B\overline{B}$ background yield is fixed to that expected from MC simulation. 
These backgrounds arise mainly from $B\to K X\gamma$ decays, where 
$X$ consists of numerous hadrons, and $B\to K^{**}\gamma$ decays, where $K^{**}$ 
denotes higher-resonance kaon states. There are small contributions from 
$B^{(0,+)}\!\to\!\rho^{(0,+)}\pi^0$ and $B^{(0,+)}\!\to\!\rho^{(0,+)}\eta^{(\prime)}$ decays.

\begin{figure*}[h]
 \includegraphics[width=58mm]{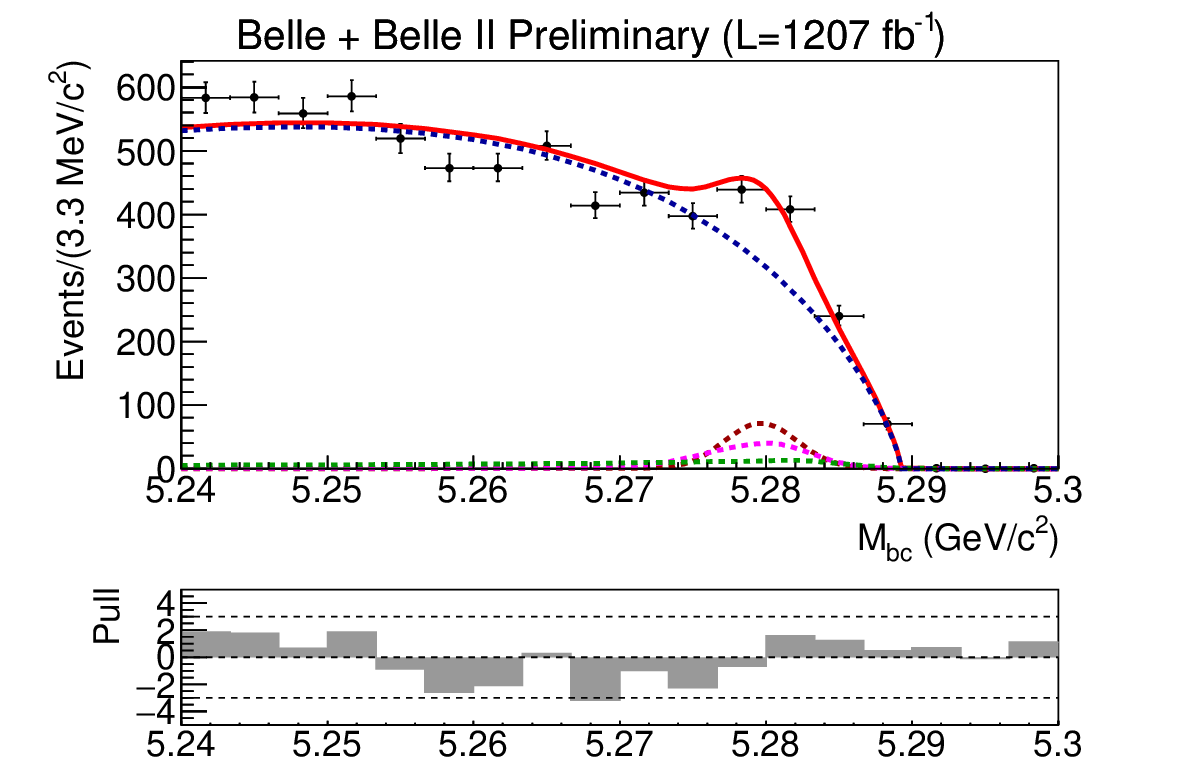}
 \includegraphics[width=58mm]{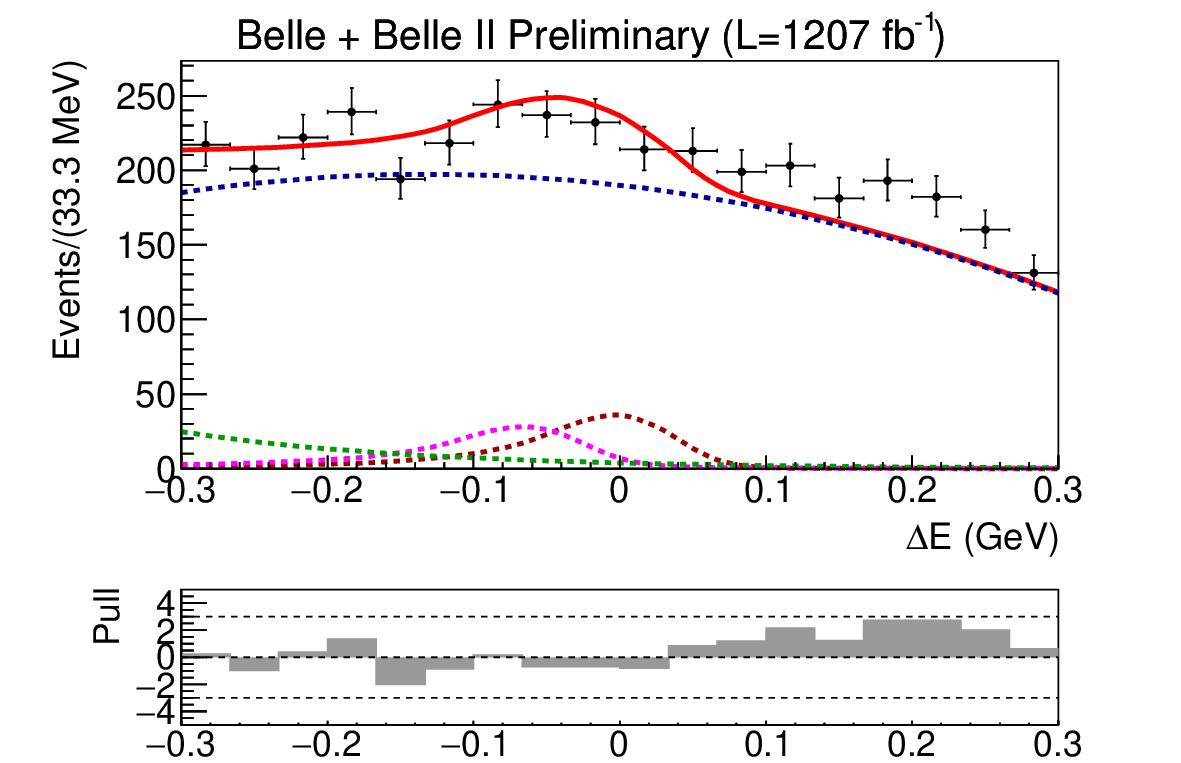}
 \includegraphics[width=58mm]{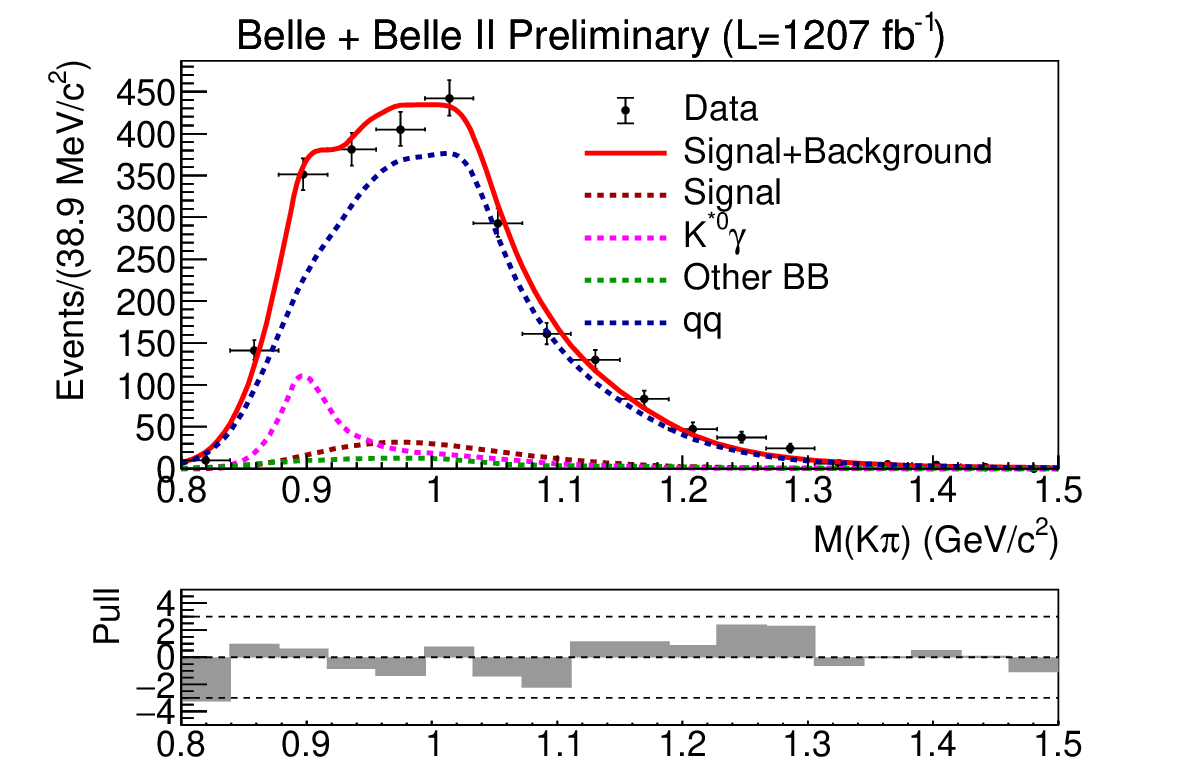}
 \caption{Distributions of $M_{\rm bc}$ (left), $\Delta E$ (middle), and $M(K\pi)$ (right), 
 combined for the three datasets, along with projections of the fit result.
 The points with error bars are data, the solid red curves are the sum of signal and background PDFs, 
 the dashed red curves are signal, the dotted-dashed blue curves are continuum background, the dashed 
 magenta curves are $K^{\ast}\gamma$ background, and the dotted green curves are $B\overline{B}$ background 
 other than $K^{\ast}\gamma$. For each of these plots, two of the three criteria 
 $M_{\rm bc}>5.27$~GeV/$c^2$, $\Delta E\in (-0.20,0.10)$~GeV, and $M(K\pi)>0.92$~GeV/$c^2$ 
 are applied, i.e., those for the variable not plotted.
 The small panels underneath the plots show the pull distributions. The deviation of the pulls 
 from zero for higher values of $\Delta E$ is due to a correlation among $\Delta E$ and $M(K\pi)$ 
 for $q\bar{q}$ events. This deviation occurs outside the $\Delta E$ range used for the $\Delta t$ 
 fit. A systematic uncertainty is evaluated for this correlation.}
 \label{fig:yield_fit}
\end{figure*}

The distributions of $M_{\rm bc}$, $\Delta E$, and $M(K\pi)$ are shown in 
Fig.~\ref{fig:yield_fit}, along with projections of the fit result. For the fit 
for the \CP-violating parameters $C$ and $S$, we select events in the 
signal-enhanced region $M_{bc}>{\rm 5.27~GeV/c^{2}}$, $\Delta E\in {\rm (-0.20,0.10)~GeV}$, 
and $M(K\pi)>0.92~{\rm GeV}/c^2$. This region is chosen to maximize the resulting 
precision for $C$ and $S$. The fitted signal yields for this region are listed in 
Table~\ref{tb:yields}.

\begin{table*}
  \caption{Signal and background yields obtained from
  the fit to $M_{\rm bc}$, $\Delta E$, and $M(K\pi)$, for the
  three datasets. The values in parentheses correspond 
  to the \CP\ fitting region (see text).}
 \begin{center}
  \begin{tabular}{lcccccccc}\hline\hline
   Dataset && $N_{\rm sig}$ && $N_{K^{\ast 0}\gamma}$ && $N_{B\overline{B}}$ && $N_{q\overline{q}}$ \\ \hline
   Belle && $84.4$ ($66.2$) && 135.4 (46.9) && 143.5 (24.0) && $5557.4$ ($639.8$) \\
   Belle~II with PXD && $68.0$ ($56.3$) && 115.8 (47.8) && 136.3 (24.9) && $7069.4$ ($879.2$) \\
   Belle~II without PXD && $14.3$ ($11.7$) && 25.9 (10.8) && 36.5 (5.2) && $1679.8$ ($201.9$) \\ 
   Total & \hspace*{0.12in} & $167\pm 23$ ($134\pm 19$) & \hspace*{0.12in} & 
           277 (106) & \hspace*{0.12in} & 316 (54) 
                         & \hspace*{0.12in} & $14307\pm 124$ ($1721\pm 15$) 
\\ \hline\hline
  \end{tabular}
  \label{tb:yields}
 \end{center}
\end{table*}

\section{Measurements of \CP\ asymmetries}
\label{sec:cp_fit}

To determine \CP-violating parameters $C$ and $S$, we
perform an unbinned maximum likelihood fit to the $\Delta t$ distributions.
The difference in decay times $\Delta t$ is calculated as
$\Delta z/\gamma\beta c$, where $\Delta z$ is the distance in $z$ between 
the decay vertex of the tag-side $B$ and that of the 
signal-side $B^0$, $\gamma\beta$ is the Lorentz factor connecting 
the $e^+e^-$ CM frame and the laboratory frame, and $c$ is the speed of light.
The number of events for each component is fixed to the yield obtained from the 
previous fit, as listed in Table~\ref{tb:yields}. 
To determine whether a signal $B$ originated as $B^0$ or $\overline{B}{}^0$, 
we implement a flavor-tagging algorithm~\cite{Belle-II:2024lwr}. 
This algorithm uses information from the non-signal $B$ decay in an event and 
outputs two quantities: $q=\pm 1$, which indicates whether the signal decay
is from $B^0$ or $\overline{B}{}^0$, and a quality factor $r$, which varies 
from 0 for no flavor information to 1 for unambiguous flavor assignment.
The event samples are divided into seven bins of $r$, 
with boundaries at 0, 0.10, 0.25, 0.45, 0.60, 0.725, 0.875, and~1.
These subsamples are fitted simultaneously for parameters $S$ and~$C$. 

The signal component is modeled by the following PDF:
\begin{eqnarray}
P^{}_k\left(\Delta t,q\,|\sigma_{\Delta t}\right) & \!=\! & 
\int \mathcal{P}^{}_{k}\!\left(\Delta t^{\prime},q\right)
\mathcal{R}\!\left(\Delta t-\Delta t^{\prime}|\sigma_{\Delta t}\right)d\Delta t^{\prime}  \nonumber \\
 & & \\
 \mathcal{P}^{}_{k}(\Delta t,q) & \!=\! & 
\frac{\exp(-\frac{|\Delta t|}{\tau_{B^{0}}})}{4\tau_{B^{0}}}\,\biggl\{1-q\Delta w^{}_k 
\nonumber \\
 & & +\ q\,a^{k}_{\varepsilon}\left(1-2w^{}_k\right) + \Bigl[q\left(1-2w^{}_k\right) \nonumber \\
 & & \hskip0.20in +\ a^{k}_{\varepsilon}\left(1-q\Delta w^{}_k\right)\Bigr] 
A_{\it CP}\!\left(\Delta t\right) \biggr\} 
\label{eq:psig}
\end{eqnarray}
where $q$ is the $B$ flavor tag;
$\mathcal{R}$ is a resolution function as defined in Ref.~\cite{Belle-II:2023bps};
$\sigma_{\Delta t}$ is the uncertainty on $\Delta t$ as determined event-by-event;
$w^{}_k$ is the wrong-tag fraction corresponding to the $r$-bin $k$ for an event; 
$\Delta w^{}_k$ is the difference between wrong-tag fractions for $q\!=\!+1$ and $q\!=\!-1$ 
for the $r$-bin $k$; 
$a^{k}_{\varepsilon}$ is the asymmetry in tagging efficiencies for $q\!=\!+1$ and $q\!=\!-1$ 
for $r$-bin $k$;
$\tau_{B^{0}}$ is the $B^0$ lifetime; and 
$A_{\it CP}\!\left(\Delta t\right)$ is defined in Eq.~(\ref{eqn:ACP}).

The distribution of signal events among the $r$ bins, and the 
flavor-tagging parameters $w$, $\Delta w$, and $a^{\rm tag}_{\varepsilon}$,
are determined from studies of 
$B^{0}\rightarrow J/\psi K^{\ast 0}$ (Belle) 
and $B^0\rightarrow D^{\ast -}\pi^+$ (Belle~II)
decays~\cite{Belle-II:2024lwr}. The uncertainties in these values are 
propagated and included as systematic uncertainties.

The parameters of $\mathcal{R}$ are determined from a $B^{0}\rightarrow J/\psi K^{\ast 0}$ 
control sample in which the $J/\psi\rightarrow\ell^{+}\ell^{-} (\ell =\mu, e)$ decay 
is ignored when fitting for the $B^0$ decay vertex. We test this procedure by fitting 
a separate sample of $B^{0}\rightarrow J/\psi K^{\ast 0}$ decays in which the 
resolution function parameters are fixed to the values obtained from the control 
sample and only the $B^0$ lifetime is floated. The resulting values of $\tau_{B^{0}}$ 
are consistent with the world average value~\cite{ParticleDataGroup:2024cfk}.

The $B^{0}\rightarrow K^{\ast 0}\gamma$ component is modeled by a similar PDF
as used for signal, with the parameter $C$ set to the world average 
value~\cite{ParticleDataGroup:2024cfk}. 
Because the final state $K^{\ast 0}$ or $\overline{K}{}^{\ast 0}$
is specific to the initial state $B^0$ or $\overline{B}{}^0$, $S$ is set to zero. 
For both $B^{0}\rightarrow\rho^{0}\gamma$ and $B^{0}\rightarrow K^{\ast 0}\gamma$ PDFs, 
$\tau_{B^{0}}$ is fixed to the $B^0$ lifetime~\cite{ParticleDataGroup:2024cfk}.

The $B\overline{B}$ background component is modeled by an exponential function
in $|\Delta t|$ convolved with the resolution function~$\mathcal{R}$. As 
$B\overline{B}$ background consists mostly of misreconstructed $B$ decays, 
an effective lifetime is used for the exponential function, as determined 
from MC simulation.

The $\Delta t$ distribution of $q\bar{q}$ background is modeled by the 
sum of two Gaussians. As done for signal events, the event-by-event
uncertainty $\sigma_{\Delta t}$ is used as a conditional variable for the 
means and widths. The parameters of the PDF are determined by fitting 
events in the $BDT_{q\overline{q}}$ sideband region.

\begin{figure}[h]
 \includegraphics[width=85mm]{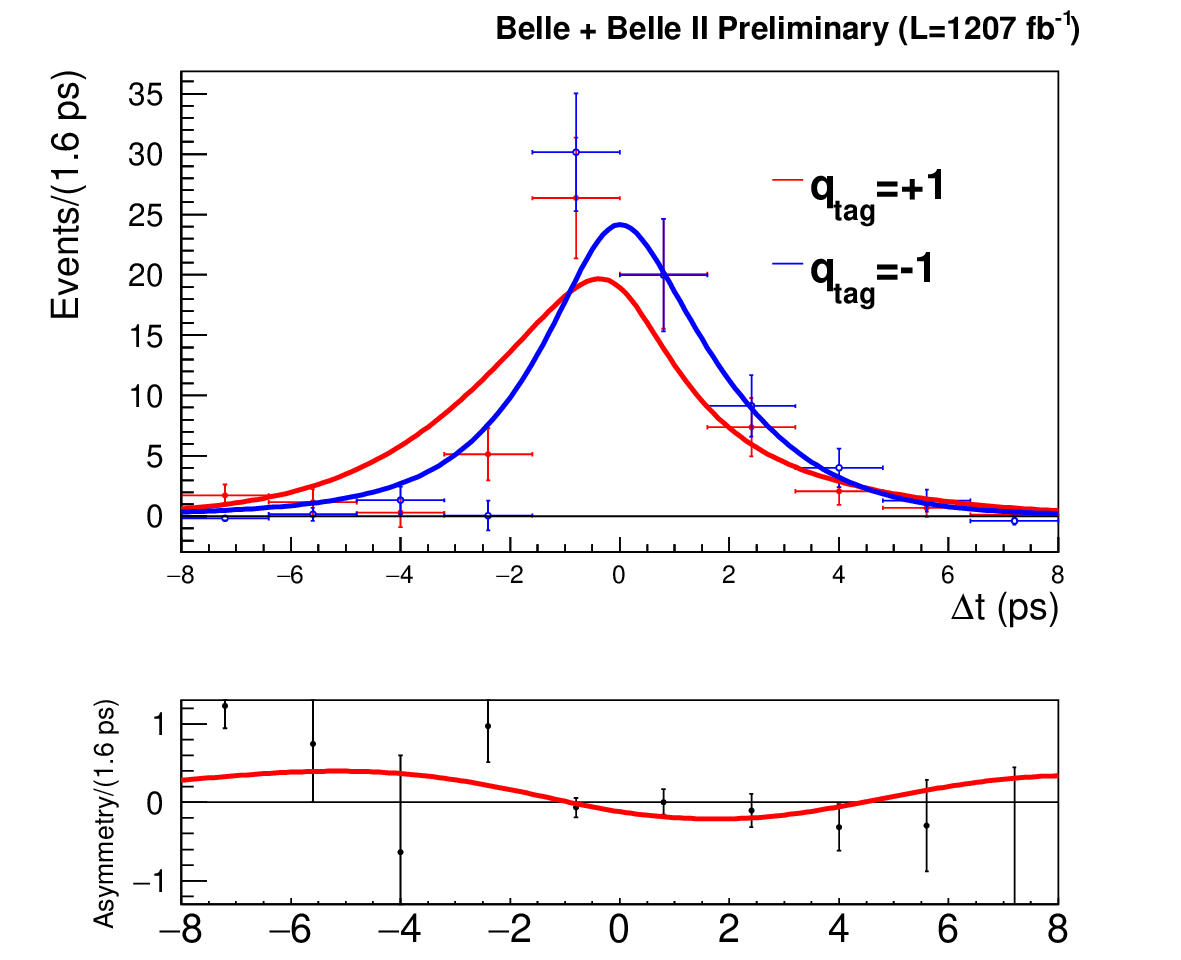}
 \caption{Top: the $\Delta t$ distributions for signal $\overline{B}{}^0$ decays (red) and 
signal $B^0$ decays (blue) obtained using {\it sPlot}~\cite{Pivk:2004ty}, along with projections 
of the fit result. All $r$ bins and datasets are summed together. Bottom: the \CP\ asymmetry 
resulting from the fit.}
\label{fig:dt_fit}
\end{figure}

The fitted $\Delta t$ distributions and projections of the fit result are 
shown in Fig.~\ref{fig:dt_fit}. The central values obtained are slightly 
outside the physical region $\sqrt{C^2 + S^2}\leq 1$. In such cases, 
we find from MC studies that the uncertainties returned by the fitter 
can be biased; i.e., the pull distributions for an ensemble of 
experiments are non-Gaussian. This is related to small signal 
samples: when the signal sample size becomes large,
this behavior disappears. To avoid this problem for our data, 
in which the signal sample size is small, we apply the following 
procedure. We map out the log-likelihood function in the $(C, S)$ 
plane and fit this distribution to a quadratic function
\begin{equation}
f(C,S)\ =\ f_{0}+\frac{1}{2}\bigl(C-C^{}_0,S-S^{}_0\bigr)\left(\begin{array}{cc} a & b \\ b & d \end{array}\right)
\left(\begin{array}{c} C-C^{}_0 \\ S-S^{}_0 \end{array}\right) . 
\end{equation} 
The central values are taken to be $C^{}_0$ and $S^{}_0$, and their uncertainties 
$\delta C$ and $\delta S$ are taken to be $\sqrt{d/(ad-b^{2})}$ and 
$\sqrt{a/(ad-b^{2})}$, respectively. 
There are points outside the physical region in which the 
PDF becomes negative and the log-likelihood cannot be calculated. For such
points, we extrapolate the log-likelihood function assuming quadratic behavior 
and using points where the log-likelihood function is calculated.
We have tested this procedure on MC simulated datasets and find good agreement
between $(C^{}_0, S^{}_0)$ and the central values returned by the $\Delta t$ fit. 
The small differences observed are assigned as systematic uncertainties.
The resulting pull distributions are Gaussian, and for $\Delta t$ fits that converge 
within 
the boundary $\sqrt{C^2 + S^2}=1$, $\delta C$ and $\delta S$ match well with the 
uncertainties from the $\Delta t$ fit. Following this procedure, we obtain 
$C^{}_{\rho\gamma} = 0.587\pm 0.230$ and $S^{}_{\rho\gamma} = -0.861\pm 0.507$, 
where the uncertainties are statistical only.

\section{Systematic Uncertainties}
\label{sec:systematics}

The various sources of systematic uncertainty are listed 
in Table~\ref{tb:syst} and estimated as discussed below. 
The total systematic uncertainty is taken as the sum in 
quadrature of all the individual contributions.

We evaluate the uncertainty arising
from the resolution function by varying all fixed 
parameters according to their uncertainties as obtained 
when determining them from the $B^{0}\to J/\psi K^{\ast 0}$ 
control sample with the $J/\psi$ ignored.
We perform the fit for $S$ and $C$ 1000 times, with each fit sampling 
different values of the resolution function parameters.
The difference between the fit result and the nominal 
result is recorded, and the standard deviation of the 
distribution of differences is taken as the systematic 
uncertainty. 

The uncertainty due to flavor tagging is evaluated in a 
similar manner. We vary the fixed parameters 
$w^{}_k$, $\Delta w^{}_k$, and $a^{k}_{\varepsilon}$,
and the fraction of decays in each $r$ bin, according to
their uncertainties~\cite{Belle-II:2024lwr} and refit the 
data. The process is repeated 1000 times, and the standard 
deviation of the distribution of differences with the 
nominal result is taken as the systematic uncertainty. 


\begin{table}[ht]
  \caption[]{Systematic uncertainties for the \CP\ measurement.
  The total systematic uncertainty is the sum in quadrature of 
  all the individual contributions.}
\renewcommand{\arraystretch}{1.3}
 \begin{center}
  \begin{tabular}{l c c}\hline\hline
   Source & $\sigma^{}_{C}$ & $\sigma^{}_{S}$ \\ \hline
   Resolution function & 0.010 & 0.048 \\
   Flavor tagging parameters & 0.029 & 0.019 \\
   Signal extraction fit & 0.022 & 0.056 \\
   $q\overline{q}$ shapes & 0.005 & 0.005 \\
   $A_{CP}$ for background & 0.010 & 0.088 \\
   Tag-side interference & 0.031 & 0.022 \\
   $\tau_{B\overline{B}}$ parameter & 0.006 & 0.004 \\
   $\tau_{B^0}$ ($B^0$ lifetime)  & 0.001 & 0.009 \\
   Log-likelihood extrapolation & 0.014 & 0.009 \\ \hline
   Total & 0.053 & 0.119 \\ \hline\hline
  \end{tabular}
  \label{tb:syst}
 \end{center}
\end{table}

To evaluate the uncertainty due to the fixed signal and 
background yields, we fluctuate the fractions of $B^{0}\rightarrow K^{\ast 0}\gamma$, 
other $B\overline{B}$ background, and $q\overline{q}$ background by their uncertainties.
For the $K^{\ast}\gamma$ background fraction, the uncertainty in
the misidentification rate $F^{}_{K\pi}$ 
is evaluated using a $B^0\to D^{*-}\pi^+$ control sample,
where $D^{*-}\to \overline{D}{}^0\pi^-$ and 
$\overline{D}{}^0\to K^+\pi^-$, and the uncertainty in 
the $\varepsilon^{}_{K^{\ast}\gamma}$ correction factor 
is evaluated from the $B^0\to K^{*0}\gamma$ control sample.
The overall uncertainty is 9.2\% (12.5\%) for Belle (Belle~II) data.
For the $B\overline{B}$ background fraction, the uncertainty is calculated 
from the uncertainties on the branching fractions of decay channels with large
contributions~\cite{ParticleDataGroup:2024cfk}. 
For those modes that have not yet been measured, we assign an uncertainty 
of 40\%, which is the largest uncertainty for measured modes. Consequently, 
the yield of $B\overline{B}$ background is fluctuated by 
13.1\% (15.7\%) for Belle (Belle~II) data. 

The uncertainty in the yield of $q\overline{q}$ background has 
two contributions, which are added in quadrature: one from the 
uncertainty in $N^{}_{q\bar{q}}$ resulting from the $M^{}_{bc}$-$\Delta E$-$M(K\pi)$ 
fit, and one due to a correlation among these observables that is neglected in the fit. 
The effect of this correlation is visible in the $\Delta E$ projection of Fig.~\ref{fig:yield_fit}
(the effect is outside the $\Delta E$ range used in the $\Delta t$ fit). To evaluate the 
latter uncertainty, we fit a large ensemble of MC-simulated samples and compare the 
resulting $N^{}_{q\bar{q}}$ yields with the input value. The largest average deviation
observed from among the three data sets (5\%) is taken as the contribution of the
correlation to the uncertainty in~$N^{}_{q\bar{q}}$.

The fixed parameters of the $q\overline{q}$ background shape 
in the fit for $C$ and $S$ are obtained from a $BDT_{q\overline{q}}$ sideband region.
These parameters are fluctuated by their uncertainties to 
estimate the uncertainty arising from the $q\bar{q}$ modeling.
The uncertainty arising from possible mis-modeling 
of the $q\overline{q}$ $M(K\pi)$ distribution is estimated as follows.
We generate 1000 MC samples according to a simplified
simulation and fit them using the nominal PDF based on kernel 
density estimation obtained from the $BDT_{q\overline{q}}$ sideband. 
The difference between the fitted signal yields and the input value 
is recorded, and the standard deviation of this distribution of differences 
is used to fluctuate the signal yield in the fit for $C$ and~$S$.

We assess the impact of a possible \CP\ asymmetry in 
$B^{0}\to K^{\ast 0}\gamma$ decays as follows. In our
nominal fit, $C^{}_{K^*\gamma}$ is fixed to the world-average 
value~\cite{ParticleDataGroup:2024cfk}, and we
fluctuate $C^{}_{K^*\gamma}$ by the corresponding
uncertainty. The parameter $S^{}_{K^*\gamma}$,
which requires interference between a decay 
amplitude resulting from mixing and a direct 
decay amplitude and thus vanishes for the $K^{*0}\gamma$ final state, 
is fixed to zero in the nominal fit. To evaluate its uncertainty, 
we fit the $B\to K^{\ast}\gamma$ control sample and 
take the value obtained (marginally consistent with 
zero) as an estimate of the possible asymmetry of
$B\to K^{\ast}\gamma$ decays in the $B^0\to\rho^0\gamma$
signal sample. We fluctuate $S^{}_{K^*\gamma}$ by this value 
($\pm 0.10$) and take the resulting shifts in $C^{}_{\rho^0\gamma}$ 
and $S^{}_{\rho^0\gamma}$ as systematic uncertainties.

For the $B\overline{B}$ background 
component, both $S^{}_{B\overline{B}}$ and $C^{}_{B\overline{B}}$ 
are fixed to zero. To evaluate a systematic uncertainty, 
we fluctuate these parameters by their respective
uncertainties. These uncertainties are calculated as the 
weighted sum of the uncertainties on $S$ and $C$ for the main decay channels 
contributing to $B\overline{B}$ background. For channels 
in which $S$ and $C$ have not yet been measured, 
we assign uncertainties of $\pm 0.10$.

The effective lifetime $\tau^{}_{B\overline{B}}$ of $B\overline{B}$ background 
is fixed in the fit to the value obtained from MC simulation. To assess
the corresponding systematic uncertainty, we vary $\tau^{}_{B\overline{B}}$ 
according to its uncertainty and repeat the fit for $S$ and $C$ 1000 times. 
The standard deviation of the distribution of differences with the nominal 
result is taken as the systematic uncertainty. 
A systematic uncertainty for $\tau^{}_{B^0}$, which is fixed to the PDG 
value~\cite{ParticleDataGroup:2024cfk}, is evaluated in a similar manner.

We assign a systematic uncertainty for the very small differences between the $(C^{}_0,S^{}_0)$ 
values obtained from our fitting procedure (fitting the two-dimensional log-likelihood 
distribution) and the central values returned by the $\Delta t$ fit. We plot the differences 
between these values for a large sample of MC-simulated events and take the widths of these 
distributions as additional systematic uncertainties.
These values are small: 1.4\% for $C_{\rho^{0}\gamma}$ and 0.9\% for $S_{\rho^{0}\gamma}$.

Finally, the effect of interference between CKM-favored and 
CKM-suppressed amplitudes for the tag-side $B$ decay is estimated as 
described in Ref.~\cite{Long:2003wq}. As a result, we assign 3.1\% 
and 2.2\% uncertainties for $C$ and $S$, respectively.

\section{Conclusion}
\label{sec:conclusion}

In summary, we have measured the time-dependent \CP\ asymmetry 
for the radiative decay $B^{0}\to\rho^{0}\gamma$ using 
1.20~$\mathrm {ab}^{-1}$ 
of data collected by the Belle and Belle~II experiments. The datasets 
together correspond to $1.29\times 10^9$ $B\overline{B}$ pairs, and 
the results supersede the previous Belle measurement based on 
$0.66\times 10^9$ $B\overline{B}$ pairs~\cite{Belle:2007kjy}. 
We obtain \CP-violating parameters 
\begin{eqnarray}
 C_{\rho^{0}\gamma} & = & \,\,0.59\pm 0.23\pm 0.05 \\
 S_{\rho^{0}\gamma} & = & \!\!-0.86\pm 0.51\pm 0.12  \,,
\end{eqnarray}
where the first uncertainty is statistical and the second is systematic. 
The correlation between the fitted values 
$C_{\rho^{0}\gamma}$ and $S_{\rho^{0}\gamma}$ is~$-0.23$.
These are the most precise measurements to date of \CP\ asymmetry parameters for $B^{0}\rightarrow\rho^{0}\gamma$.
The value for $C^{}_{\rho^0\gamma}$ is consistent within $1.7\sigma$ 
with the SM prediction of $\sim\!0.17$~\cite{Beneke:2004dp}.
The value for $S^{}_{\rho^0\gamma}$ is consistent within $1.6\sigma$ 
with the SM prediction of $<\!0.02$~\cite{Ball:2006eu}.

\section*{Acknowledgements}
This work is based on data collected using the Belle~II detector, which was built and commissioned
prior to March 2019, and the Belle detector, which was operated until June 2010. We are grateful for
support from the following funding agencies:  
the Higher Education and Science Committee of the Republic of Armenia Grant No. 23LCG-1C011; 
Australian Research Council and Research Grants No. DP200101792, No. DP210101900, No. DP210102831, No. DE220100462, 
No. LE210100098, and No. LE230100085; 
Austrian Federal Ministry of Education, Science and Research, Austrian Science Fund (FWF) 
Grants DOI: 10.55776/PAT1836324, DOI: 10.55776/PAT1823724, and DOI: 10.55776/PAT3412626; 
Natural Sciences and Engineering Research Council of Canada, Digital Research Alliance of Canada, 
and Canada Foundation for Innovation; 
National Key R\,\&\,D Program of China under Contract 2022YFA1601903, No. 2024YFA1610503, and No. 2024YFA1610504, 
National Natural Science Foundation of China and Research Grants No. 11575017, No. 11761141009, No. 11705209, 
No. 11975076, No. 12135005, No. 12150004, No. 12161141008, No. 12405099, No. 12475093, and No. 12335004, 
and Shandong Provincial Natural Science Foundation Project ZR2022JQ02; 
Czech Science Foundation Grant No. 22-18469S, Regional funds of EU/MEYS: OPJAK FORTE CZ.02.01.01/00/22 008/0004632, 
and Charles University Grant Agency project No. 246122; 
European Research Council, Seventh Framework PIEF-GA-2013-622527, Horizon 2020 ERC-Advanced Grants No. 267104 and 
No. 884719, Horizon 2020 ERC-Consolidator Grant No. 819127, Horizon 2020 Marie Sklodowska-Curie Grant 
Agreement No. 700525 “NIOBE” and No. 101026516, and Horizon Europe Marie Sklodowska-Curie Staff Exchange 
project JENNIFER3 Grant Agreement No. 101183137 (European grants), 
L’Institut National de Physique Nucl´eaire et de Physique des Particules (IN2P3) du CNRS under 
Project Identification No. CNRS-IN2P3-14-PP-033, and L’Agence Nationale de la Recherche (ANR) 
under Grant No. ANR-23-CE31-0018 and ANR-25-CE31-1333 (France); 
BMFTR, Deutsche Forschungsgemeinschaft (DFG), HGF, MPG, AvH Foundation, the Carl Zeiss Foundation, 
and the Volkswagen Foundation (Germany); 
Department of Atomic Energy under Project Identification No. RTI 4002, Department of Science and Technology, 
and UPES SEED funding programs No. UPES/R\,\&\,D-SEED-INFRA/17052023/01 and No. UPES/R\,\&\,D-SOE/20062022/06 (India); 
Israel Science Foundation Grant No. 2476/17, U.S.-Israel Binational Science Foundation Grant No. 2016113, and 
Israel Ministry of Science Grant No. 3-16543; 
Istituto Nazionale di Fisica Nucleare and the Research Grants BELLE2, and the ICSC-Centro Nazionale di Ricerca 
in High Performance Computing, Big Data and Quantum Computing funded by European Union-NextGenerationEU; 
Japan Society for the Promotion of Science, Grant-in-Aid for Scientific Research Grants No. 16H03993, 
No. 16H06492, No. 16K05323, No. 17H01133, No. 17H05405, No. 18K03621, No. 18H03710, No. 18H05226, No. 19H00682, 
No. 20H05850, No. 20H05858, No. 22H00144, No. 22K14056, No. 22K21347, No. 23H05433, No. 26220706, No. 26400255, 
and No. 26H02056, the Ministry of Education, Culture, Sports, Science, and Technology (MEXT) of Japan, 
and Tau-Lepton Physics Research Center of Nagoya University; 
National Research Foundation (NRF) of Korea Grants No. 2021R1-F1A-1064008, No. 2022R1-A2C-1003993, No. RS-2018-NR031074, 
No. RS-2021-NR060129, No. RS-2024-00354342, No. RS-2025-02219521, No. RS-2026-25471491, No. RS-2026-25480677, and 
No. RS-2026-25486791, Radiation Science Research Institute, Foreign Large-Size Research Facility Application Supporting 
Project, the Global Science Experimental Data Hub Center, the Korea Institute of Science and Technology Information (K26L1M2C3) 
and KREONET/GLORIAD; 
Universiti Malaya RU grant, Akademi Sains Malaysia, and Ministry of Education Malaysia; 
Frontiers of Science Program Contracts No. FOINS-296, No. CB-221329, No. CB-236394, No. CB-254409, 
and No. CB-180023, and SEP-CINVESTAV Research Grant No. 237 (Mexico); 
the Polish Ministry of Science and Higher Education, and the National Science Center; 
the Ministry of Science and Higher Education of the Russian Federation and the HSE University Basic Research Program, Moscow; 
University of Tabuk Research Grants No. S-0256-1438 and No. S-0280-1439 (Saudi Arabia); 
Slovenian Research Agency and Research Grants No. J1-50010 and No. P1-0135; 
Ikerbasque, Basque Foundation for Science, State Agency for Research of the Spanish Ministry of Science and 
Innovation through Grant No. PID2022-136510NB-C33, Spain, the Severo Ochoa project CEX2023-001292-S funded by MICIU/AEI, 
State Secretariat for Telecommunications and Digital Infrastructure with reference TSI-069100-2023-0012, 
State Agency for Research of the Spanish Ministry of Science, Innovation and Universities through 
Grant No PID2024-156645NB-C21; 
the Swiss National Science Foundation; 
the Knut and Alice Wallenberg Foundation (Sweden), Contracts No. 2021.0174, No. 2021.0299, and No. 2023.0315; 
National Science and Technology Council, and Ministry of Education (Taiwan); 
Thailand Center of Excellence in Physics; 
TUBITAK ULAKBIM (Turkey); 
National Research Foundation of Ukraine, Project No. 2020.02/0257, and Ministry of Education and Science of Ukraine; 
the U.S. National Science Foundation Grants No. PHY-1913789 and No. PHY-2111604, 
and the U.S. Department of Energy Awards No. DE-AC06-76RLO1830, No. DE-SC0007983, No. DE-SC0009824, 
No. DE-SC0009973, No. DE-SC0010007, No. DE-SC0010073, No. DE-SC0010118, No. DE-SC0010504, No. DE-SC0011784, 
No. DE-SC0012704, No. DE-SC0019230, No. DE-SC0021616, No. DE-SC0022350, and No. DE-SC0023470; 
and the Vietnam Academy of Science and Technology (VAST) under Grant No. DL0000.05/26-27. 

We thank the SuperKEKB and the KEKB teams for delivering high-luminosity collisions; the KEK cryogenics group for the 
efficient operation of the detector solenoid magnet and IBBelle on site; the KEK Computing Research Center for on-site 
computing support; the Pacific Northwest National Laboratory (PNNL) Environmental Molecular Sciences Laboratory (EMSL) 
computing group for strong computing support; the NII for SINET6 network support; and the raw-data centers hosted by 
BNL, DESY, GridKa, IN2P3, INFN, and the University of Victoria.

These acknowledgements are not to be interpreted as an endorsement of any statement made by any of our institutes, 
funding agencies, governments, or their representatives.

\section*{Data Availability}
\ifdefvoid{\hepdata}{}{Numerical data corresponding to the results presented are available as HEPData.}
The full Belle II data are not publicly available.
The collaboration will consider requests for access to the data that support this article.

\bibliography{belle2-references,references}

\end{document}